\documentclass[journal]{journal}
\usepackage{graphicx}
\usepackage[caption=false]{subfig}

\begin{document}

\title{Wearable Affective Memory Augmentation}
%
%
%

\author{Cayden~Pierce and
        Steve~Mann
}

\maketitle
\thispagestyle{empty}


\begin{abstract}

Human memory prioritizes the storage and recall of information that is emotionally-arousing and/or important in a process known as value-directed memory. When experiencing a stream of information (e.g. conversation, book, lecture, etc.), the individual makes conscious and subconscious value assessments of the incoming information and uses this as a metric to determine what to remember. In order to improve automatic recall of memory, previous memory augmentation systems have sensed users' physiological state to determine which sensory media should be prioritized. Here, we propose to prioritize memories using the affective state of individuals that the user is interacting with. Thereby, the proposed wearable Affective
Memory Augmentation system uses affective information from the user’s social companions in order to facilitate value-directed memory.
\end{abstract}


\IEEEpeerreviewmaketitle

\section{Introduction}

\IEEEPARstart{W}{earable} computers (wearables) can improve human cognitive capabilities by symbiotically integrating into our thinking processes. Systems that work in this way are described by Humanistic Intelligence (HI), a framework whereby the human and machine work together, as a single system, to perform cognitive functions~\cite{minsky2013society}. Memory is a fundamental component of human cognition~\cite{posner1973cognition} which allows us to learn, recognize, reason, and remember. We propose to improve human memory capabilities using an HI wearable computing system.

It has been demonstrated that memory is positively correlated with intelligence~\cite{shelton2010relationships, alexander1997intelligence, colom2007general}. Thus, improving users' memory capabilities can be utilized as an effective method to improve user intelligence. To accomplish this, memory augmentation systems are tools which aim to enhance users' memory capabilities by improving their ability to encode, store, and recall information. An ideal memory augmentation system will enhance the memory capabilities of users with normal memory functionality as well as restore basic memory functionalities to individuals with memory impairments (e.g. Alzheimer's, dementia).

Memory impairments are diseases, injuries, conditions, etc. which negatively impact human memory function~\cite{jonker1996memory}. All of an individual's personal knowledge is represented in memory and new information cannot be learned without the use of memory~\cite{lieberman2012human}. These mental necessities of knowledge and learning make memory an incredibly important component of cognition, and are part of the reason why memory impairments have such a strongly negative impact on those afflicted. Further, as we age, our memory steadily declines~\cite{grady2000changes} and we experience increased risk for severe memory diseases/impairments such as Alzheimer's, dementia, and mild cognitive impairment (MCI)~\cite{guerreiro2015age}. Methods that help deal with the symptoms of memory impairment could improve the lives of millions of people living with memory deficiencies.

\begin{figure}
\includegraphics[width=\columnwidth]{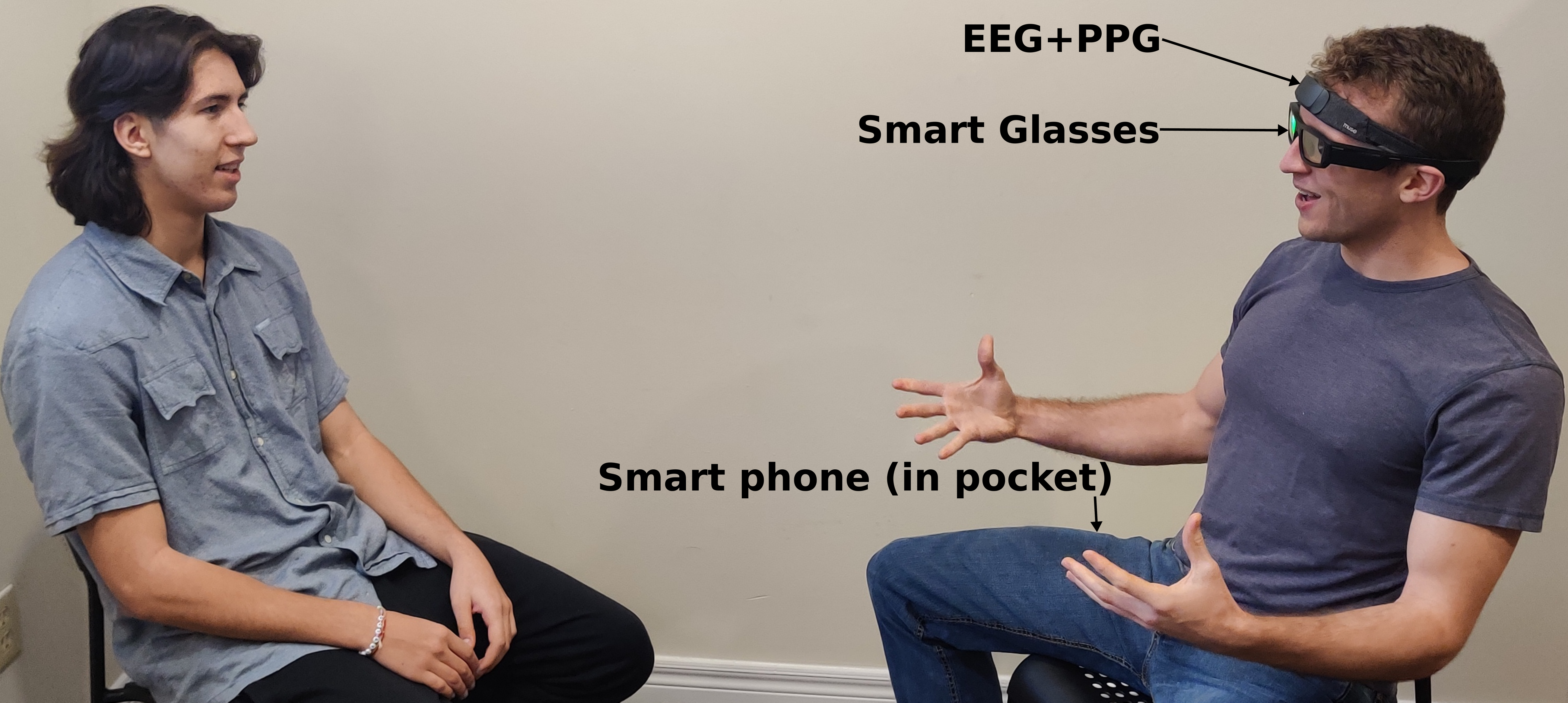}
\caption{A conversation where a user (right) utilizes the Affective~\protect\cite{picard97} Memory Augmentation system to remember the most important information. The user is shown wearing the full prototype system consisting of Vuzix Blade Upgraded smart glasses, Muse S electroencephalography (EEG) and photoplethysmography (PPG) system, and an Android 10 smart phone.
}
\label{fig:cyborg_conversation}
\end{figure}


\subsection{Priming}
Human memory is highly associative - memories relate to other memories by their conceptual, temporal, and semantic properties~\cite{foss1975memory, yonelinas2019contextual}. This differs significantly from classical computer memory, which is based upon numerical addressing schemes for data. Priming is the psychological effect whereby exposure to an initial stimulus affects an individuals response to subsequent stimuli. In the context of human memory, memory priming occurs when the mention or recall of information conceptually, temporally, and/or semantically related to a certain memory will make the recall of that memory easier, faster, and more likely~\cite{ratcliff1988retrieval}.

\subsection{Value-Directed Memory}

An aspect of human memory function is value-directed memory~\cite{stefanidi2018free, castel2012metamemory}. Value-directed memory, or value-directed remembering, occurs when human memory prioritizes the storage and recall of information that is considered emotionally-arousing~\cite{hadley2006does, lee2015encoding} and/or important by the individual. When experiencing a stream of information (e.g. conversation, book, lecture, etc.), the individual makes conscious and subconscious~\cite{mitchell2002directed} value assessments of the incoming information and uses this as a metric to determine what to remember.

\subsection{Computer Induced Flashbacks}

In early memory augmentation systems~\cite{mann1996wearable, mannvmp}, it was discovered that recording of sensory experience and subsequent presentation of a short snippet of that stimuli to a user can have a highly effective memory priming effect on the user. For example, displaying a few frames of video from a past event is often sufficient to prime the user's memory such that they immediately recall detailed memories of the presented place and time. These digitally created memory flashbacks are known as \textit{computer-induced flashbacks}.

\section{Background}

Some of the earliest invented human technology was designed for memory augmentation, such as cave art, clay tablets, and other means to record information. Memory aiding technology is still fundamental today, as pen, paper, note taking applications, personal knowledge management (PKM) computing systems, virtual whiteboards, etc. are some of the most commonly used day-to-day technologies.

Modern memory augmentation research has three major foci: intelligent storage of information (what should be stored), intelligent recall of information (what should be recalled), and human-computer interaction (how should users request information, how should information be presented, how do we present memories in a way that is a natural extension of the human thinking process, etc.).

\subsection{Previous Memory Augmentation Systems}
\subsubsection{Memex}
Early computing pioneers realized the possibility of a special-purpose memory augmentation machine designed specifically to increase the speed, efficiency, and associative capabilities of memory augmentation. The Memex system, described by Vannevar Bush~\cite{bush1945we}, envisioned a new way for man and machine to work together as a vastly extended memory system. The Memex gave the user the ability to store all of one's paper-based information in a machine that would store data, provide it on demand, and allow for linking between documents. Modern PKM systems have achieved many of the initial goals of the Memex system.

\subsubsection{Visual Memory Prosthetic}
The Visual Memory Prosthetic (VMP)~\cite{mannvmp} is a wearable computer camera system that records the user's life (lifelogging or glogging). The user is then able to re-experience their memories via the wearable interface. This re-experiencing can elicit computer-induced flashbacks, greatly improving the user's ability to remember an event. Further development of the VMP added user physiological sensing to the device. Understanding the physiological state of the user allowed the memory augmentation device to simulate human value-directed remembering by prioritizing memories that the user found interesting, engaging, and relevant. For individuals with visual memory deficiencies, this emulates how human memory prioritizes information deemed important, and thus serves as an effective aid to users with memory impairments.

\subsubsection{Remembrance Agent}
The Remembrance Agent~\cite{Starner93} is a memory augmentation system which stores information that is consciously input by the user through a wearable keyboard (e.g. chording keyboard). As the user takes notes, the most recently recorded data is used as contextual information to search through a user's textual databases of notes, email, memos, etc. to recall and display relevant information to the current task, thought, or conversation.

\subsubsection{NeverMind}
The NeverMind system~\cite{rosello2016nevermind} employs the memory-palace/method-of-loci memory technique~\cite{legge2012building} to help users remember sequential information. Classical memory-palace techniques involve the individual using their spatial memory to connect items in a list to spatial positions in a room, house, or route. The individual can recall the list by mentally traversing through that space. NeverMind automates this method by providing an augmented-reality (AR) interface with sequential items placed directly in 3D space in front of the user. This method significantly improves users' ability to remember/memorize sequential lists of information, and explores new methods of interface for memory.



\begin{figure}
\centering
\includegraphics[height=2.6in]{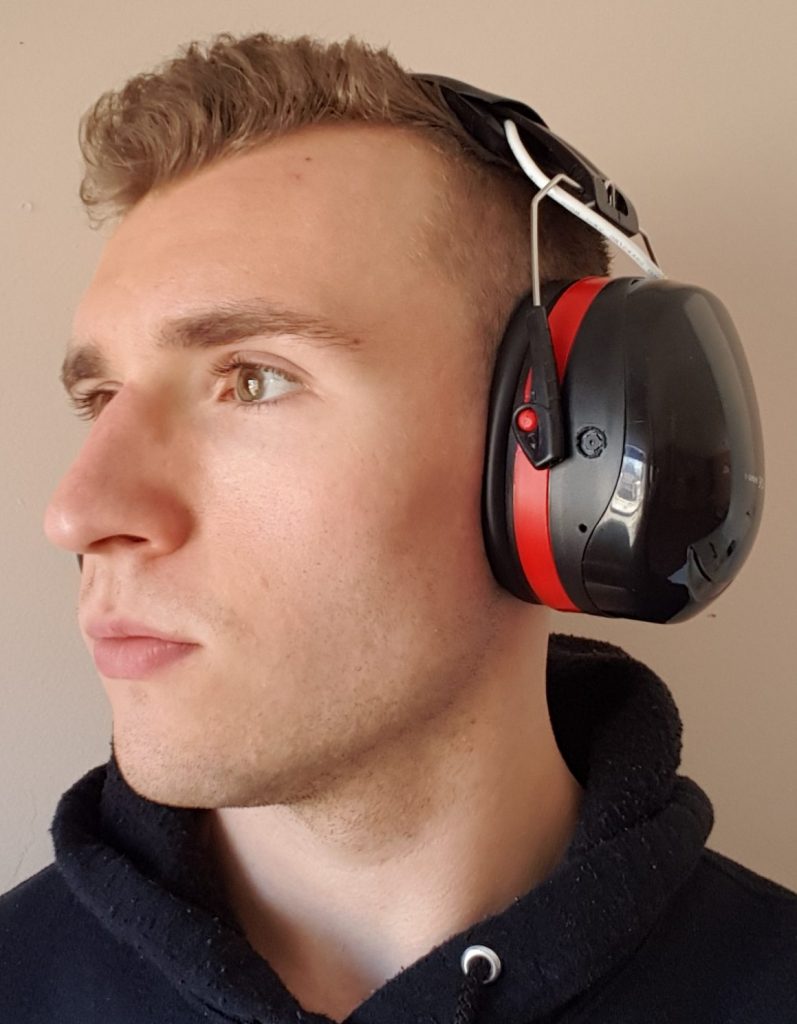}
\caption{Early headphones form factor version of the wearable Affective Memory Augmentation system. This prototype is running simultaneous (and independent) memory augmentation and affective computing tools.
}
\label{fig:emex_vmp}
\end{figure}

\begin{figure}
\includegraphics[width=\columnwidth]{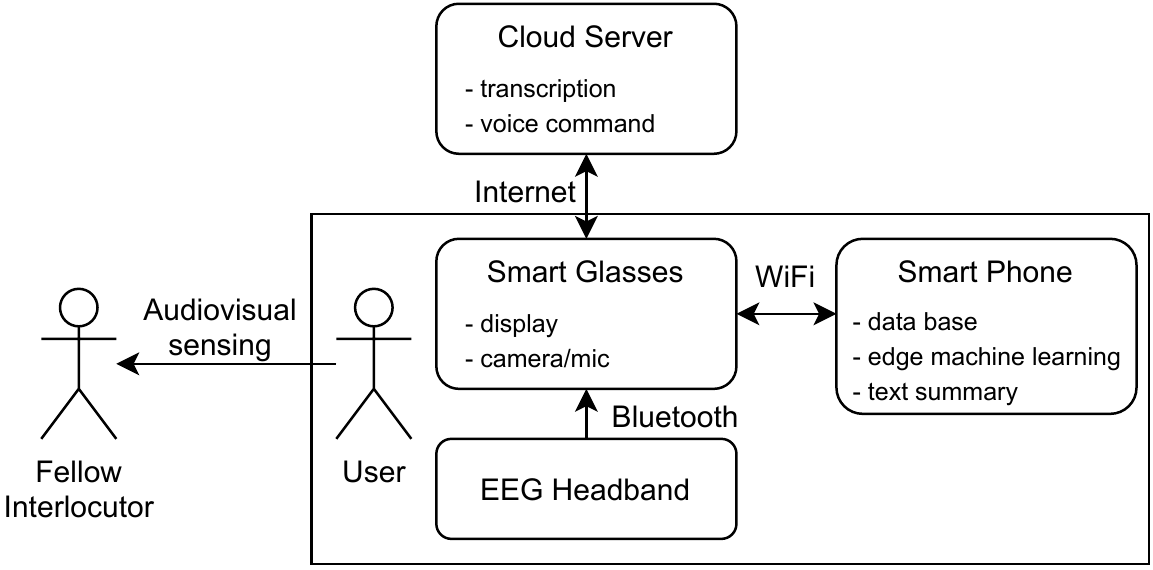}
\caption{
A system diagram of the Affective Memory Augmentation system.
}
\label{fig:system_diagram}
\end{figure}

\section{Wearable Affective Memory Augmentation System}

Value-directed memory is a function of human memory whereby perceptually important and emotionally arousing events are remembered better than less important events. This is individualistic, whereby the individual makes a personal, subconscious appraisal of the value of some information in order to prioritize its encoding, storage, and recall. However, in social situations, it is not ideal for an individual to only remember what they personally thought was important. It is just as vital for the individual to remember what their fellow interlocutors deemed important. Thus, the Affective Memory Augmentation system uses affective information~\cite{picard95, picard97a} from the user's social companions in order to facilitate value-directed memory.

This work builds on previous memory augmentation systems~\cite{Starner93, rosello2016nevermind, bahrainian2018augmentation,greenwald2015taketwo, devaul2004memory} by improving the systems' ability to prioritize information according to its value/importance. Some previous memory augmentation systems have used the user's physiological state (as recorded by wearable sensors) as the guiding metric to perform value-directed remembering. These memory augmentation systems perform automated value-directed memory, whereby memories are weighted as more or less important depending on how the user's state changed when the information was received.

We propose to extend this technique by not only measuring the physiological state of the wearer, but also of other people that the user is interacting with. Each individual in a social situation has their own appraisal of the value of the current information/experience, which they communicate to others via verbal and non-verbal affective cues. These social communicative cues can be sensed with wearable, egocentric sensors and analyzed using Affective Computing signal processing techniques. The combination of value appraisal from multiple individual sources will improve the accuracy of value-directed remembering in memory augmentation systems.

Finally, this new method allows for affectively-informed automatic summarization of conversations, affective search through one's extended memory, extraction of the most salient information from a social gathering, and other novel memory augmentation use cases made possible by an Affective Computing memory augmentation system.

\subsection{Early Prototype}

An early prototype of the Affective Memory Augmentation system utilizes a custom built wearable computer in headphones form factor. This early iteration simultaneously promotes the memory and affective intelligence of the user. Memory is promoted via a wearable face recognizer using point-of-view face recognition to recognize peoples' faces and help the user remember their names. The system simultaneously runs human pose estimation to process non-verbal communication of identified individuals, extracting stress and confidence scores from body language. This social memory and affective information is presented to the user via audio feedback. The early prototype system is shown in Fig.~\ref{fig:emex_vmp}.

\subsection{Current System Overview}
The current prototype of the Affective Memory Augmentation system has been updated to a smart glasses form factor. An image of the system in use can be seen in Fig.~\ref{fig:cyborg_conversation}. A system overview diagram can be viewed in Fig.~\ref{fig:system_diagram}. Smart glasses (Vuzix Blade Upgraded) are worn by a user in a social, conversational setting. Optionally, user's can wear a mobile electroencephalography (EEG) and photoplethysmography (PPG) headband (Muse S by InterAxon) to measure insights into their own physiological state.

The smart glasses continuously capture point-of-view (POV) video, POV audio, EEG, PPG, and other sensory streams (inertial measurement unit, global positioning system (GPS), etc.). This data is wirelessly sent to an Android smart phone over WiFi and WebSocket protocols. The Android smart phone is running a signal processing pipeline (a MediaPipe~\cite{lugaresi2019mediapipe} Perception Pipeline designed for WearableAI~\cite{wearableai} applications). This Perception Pipeline runs frames of the video through a number of Affective Computing neural networks (facial emotion, eye contact, pose estimation, etc.) and records streams of both raw data and affective information to an Android Room database.

The smart glasses also stream raw, encrypted audio to a cloud server (AWS EC2 Ubuntu Linux running Python) to run automatic speech recognition (ASR) and parse user voice commands. The voice command system listens for voice commands and annotations (e.g. a user command to signal the start of a new conversation). Resultant transcriptions are returned, displayed on the smart glasses display, and saved locally in the Android Room database.

Whenever the user wants to search through their memory, create a conversation summary, induce a computer-induced flashback of an event, etc., they may run the function by speaking a voice command or by utilizing a user interface (UI) on the Android smart phone.

\subsection{Most Salient Raw Information}

Like an automated highlight reel of one's social interactions, this memory augmentation functionality extracts the most salient moments from a period of time and presents them to the user. This functionality uses a combination of physiological information about the user and affective information from the user's social companions to locate the most salient moments from a period of time. Transcriptions (or, optionally, audio and video) of these peaks in salience are then extracted and displayed to the user.

To do so, the physiological and affective data streams are first processed for indications of salience. For example, increased heart rate and respiratory rate serves as a physiological indication of salience, while excited facial expression, high levels of eye contact, and open body language are affective indicators of salience. If users choose not to utilize the EEG+PPG headband, then the processing is run on only the affective data. Time series data is processed in a sliding window to create a time series representation of the salience of social interactions. This data is then searched to find the n timestamps of highest salience. The system then extracts conversation transcription snippets nearest to those times of high salience. Finally, the resulting transcription snippets are sent from the smart phone to the smart glasses to be displayed in an ordered list on the user's smart glasses display. 

\subsection{Affective Conversation Summary}

\begin{figure}
\includegraphics[width=\columnwidth]{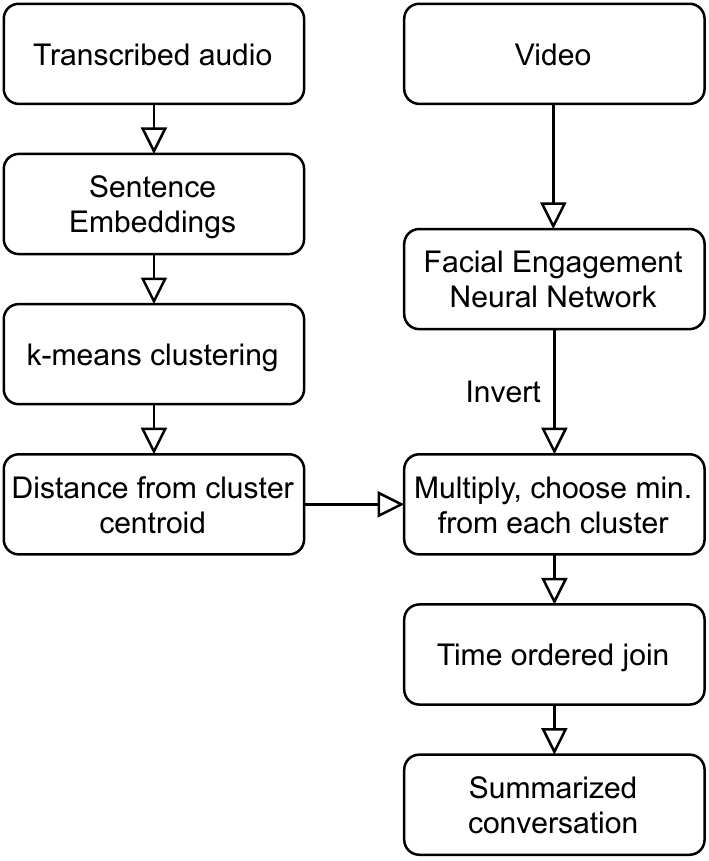}
\caption{Signal processing pipeline using Wearable Affective Computing to extract affective summaries of conversations.
}
\label{fig:affective_summary}
\end{figure}

Priming can be utilized to promote memory recall. When sufficient priming cues are presented to the user, a computer induced flashback is induced. We attempt to increase the efficacy of memory priming cues in eliciting computer induced flashbacks by presented summaries of memories whereby the summarization is generated using affective information~\cite{singhal2018summarization}. Since important information and affectively-charged information are better remembered and recalled, using this information as memory cues should result in more effective priming. 

First, a user marks the beginning and end of a specific conversation or social interaction using voice commands. A list of these user-annotated events are then stored on the user's smart phone. Through a smart phone user interface displaying a list of these events, users may make a selection of a specific event, prompting the system to run the Affective Conversation Summary on that event. The result of the summary is then presented as a text paragraph on the user's smart glasses display.

Textual summarization is the process of compressing a large body of natural language text into a shorter natural language representation. Effective summarization methods significantly reduce the size of the text whilst maintaining the most important information. Common approaches to textual summary include \textit{extractive summarization}, which chooses the n most important sentences from the text and joins them together as the summary. \textit{Abstractive summarization} also extracts and prioritizes information, but in an embedded semantic representation, and uses this semantic representation to create an entirely new piece of text which briefly describes the input text.

We use state of the art extractive summarization techniques combined with contextual affective information to create a summary of a social interaction which is informed by the engagement of the participants. If the user and/or their conversation partner is highly engaged and absorbed at a particular instance, the transcript at that point in time will be up-weighed and thus more likely to appear in the output summary text. Similarly, points in conversation of less interest will be down-weighed and thus less likely to appear in the output summary text.

The affective information used to prioritize the text is engagement. We utilize facial emotion recognition and facial engagement machine learning models~\cite{DBLP:journals/corr/abs-1808-02324} to extract engagement information from video.

We utilize state of the art techniques in natural language processing (NLP) for extractive text summarization, as described by~\cite{padmakumar2016unsupervised, khan2020sentence}. These techniques were extended with state of the art sentence embedding transformers~\cite{hugface_minilm} and a simple method of combining the text summarization technique with Affective Computing insights. The processing pipeline can be viewed in Fig.~\ref{fig:affective_summary} and is described below:
\begin{enumerate}
\item Separate input text into a list of sentences
\item Process affective information regarding engagement at each sentence timestamp, tag the sentence between 0 (not engaged) and 1 (fully engaged)
\item Generate sentence embeddings using a transformer model 
\item Cluster sentence embeddings into n clusters, (where n is equal to the number of sentences in our output summary)
\item For each sentence embedding, compute distance from centroid (mean) of cluster
\item Multiply the centroid distance of each sentence embedding by the \textit{inverted} affective engagement score ($invertedAffectiveScore = 1 - affectiveScore$) to create an updated distance score, the lower this updated distance score, the more important the sentence
\item From each cluster, choose the sentence with the lowest updated distance score
\item Join chosen sentences based on the order they originally appeared in the input text
\end{enumerate}

The resulting affective conversation summary is displayed to the user on the smart glasses display.

\subsection{Affective Memory Search}

Constantly recording and extracting affective information from audio, video, transcriptions, and various sensors results in a multi-dimensional time-series affective signal of the user's life. This presents a unique opportunity to search through one's life, not by time or place, but by the affective context of one's experiences. For example, one may wish to explore their own memory with the following questions: 

\begin{itemize}
\item When was my partner happiest in the last week?
\item Where do I experience the most stress?
\item What did I say that caused Jane's excitement in our conversation on Friday?
\item Which conversation topic caused my daughter the most stress?
\end{itemize}

As an initial exploration of this possibility, we implement a voice command Affective Memory Search system that allows users to search for the peak of a specific emotion during a social interaction. For example, the user might say \textit{``hey computer, affective search happiness"}. This will find the happiest moment in the user's previous conversation (as understood by facial emotion recognition on POV camera feed), extract the transcription, audio, or video from that point in time, and display it to the user. The user can search for any facial emotion with this method, including sadness, happiness, fear, disgust, anger, surprise, and neutrality.

\section{Future Work}
Earlier, we explored how a number of functional systems have been built and demonstrated to positively augment memory. Many of these approaches could be greatly improved in terms of memory predictive recall as well as human-computer interaction by two major developments. 

First, since the development of the previous systems, improvements in signal processing techniques from noisy, complex, high dimensional data has been greatly improved by machine learning and deep learning techniques. Processing of egocentric data, both to understand the content (object recognition, scene recognition, facial recognition, automatic speech recognition, etc.) and the context (biosensory processing, affective computing processing, etc.) has been significantly improved. Thus, the accuracy and therefore usefulness of previous memory augmentation systems would be enhanced if updated with these latest techniques. Further, associative memory systems~\cite{Starner93} discussed could be improved by updating their associative inference systems to deep learning approaches based on multi-modal data and multi-modal deep neural networks. 

Another significant development has been in wearable hardware, whereby the wearable computers developed now are smaller, more comfortable, and thus more realistic for all day and everyday use. This is important beyond simply a consumer or industry necessity - it is fundamental to the research because only when a memory augmentation system is used daily and consistently does it reveal its potential, strengths, and weaknesses.

Future work will focus on combining affective memory augmentation (presented herein), previous memory augmentation systems, and state of the art signal processing and wearable hardware developments in to a single, cohesive memory enhancement system.

\section{Conclusion}

We proposed a wearable computing system which uses Affective Computing insights to augment and enhance human memory. The process of value-directed memory was guided by physiological and affective metrics within a social interaction in order for the memory augmentation system to understand and prioritize information that is most important for the user to remember. We then developed and tested memory tools which utilize this affective information to summarize conversations, extract moments of greatest salience in social interactions, and search through time for peaks of specific emotions.


\appendices

\section*{Acknowledgment}

The authors would like to thank Vuzix for donation of Vuzix Blade smart glasses and InterAxon for donation of Muse mobile EEG+PPG systems. We would also like to thank Alexander for being a model (photographic subject) in Fig.~\ref{fig:cyborg_conversation}
and Callum for photographing Fig.~\ref{fig:cyborg_conversation}.

\ifCLASSOPTIONcaptionsoff
  \newpage
\fi

\IEEEtriggeratref{31}
\bibliographystyle{IEEEtran}
\bibliography{chirplet}

\end{document}